\documentclass[a4paper]{jpconf}
\usepackage{graphicx}
\usepackage{subcaption}
\usepackage{here}
\usepackage{amssymb}
\begin{document}
\title{Light charged Higgs boson in $H^\pm h$ associated
	production at the LHC}

\author{A. Arhrib$^1$, R. Benbrik$^2$, M. Krab$^3$, B. Manaut$^3$, S. Moretti$^{4,5}$, Y. Wang$^6$ and Q.-S. Yan$^{7,8}$}

\address{$^1$ Abdelmalek Essaadi University, Faculty of Sciences and Techniques,
	Tangier, Morocco}
\address{$^2$ Laboratory of Fundamental and Applied Physics, Polidisciplinary, Sidi Bouzid,
	BP 4162, Safi, Morocco}
\address{$^3$ Research Laboratory in Physics and Engineering Sciences, Modern and Applied Physics Team, Polidisciplinary Faculty, Beni Mellal, 23000, Morocco}

\address{$^4$ School of Physics and Astronomy, University of Southampton,
Southampton SO17 1BJ, U.K.}

\address{$^5$ Department of Physics and Astronomy, Uppsala University, Box 516, SE-751 20 Uppsala, Sweden}

\address{$^6$ College of Physics and Electronic Information, Inner Mongolia Normal University,
Hohhot 010022, P.R. China}

\address{$^7$ Center for Future High Energy Physics, Chinese Academy of Sciences,
Beijing 100049, P.R. China}

\address{$^8$ School of Physics Sciences, University of Chinese Academy of Sciences,
Beijing 100039, P.R. China}

\ead{aarhrib@gmail.com, r.benbrik@uca.ma,
	mohamed.krab@usms.ac.ma, b.manaut@usms.ma, s.moretti@soton.ac.uk; stefano.moretti@physics.uu.se, wangyan@imnu.edu.cn, 
	yanqishu@ucas.ac.cn}

\begin{abstract}
We investigate the light charged Higgs boson production via $pp \rightarrow H^\pm h$ at the Large Hadron Collider (LHC) in the Two-Higgs Doublet Model (2HDM) type-I. By focusing on a scenario where $H$ is the observed Higgs boson of mass 125 GeV, we study the aforementioned Higgs boson production channel and explore its bosonic decay $H^\pm \rightarrow W^\pm h$. The latter can have a sizable rate that possibly dominates, over the fermionic decay channels, in several regions of the viable parameter space. We demonstrate that the production process $pp \rightarrow H^\pm h$ followed by $H^\pm \rightarrow W^\pm h$ could be the most promising experimental avenue to search for light $H^\pm$ at the LHC.
\end{abstract}

\section{Introduction}
The Standard Model (SM) is the most successful framework that describes the physics of elementary particles. In 2012, a Higgs boson was discovered at the Large Hadron Collider (LHC) \cite{ATLAS:2012yve, CMS:2012qbp}, which is the only required puzzle piece to confirm, in particular, the mechanism of ElectroWeak Symmetry Breaking (EWSB) and thereby the SM. However, the origin of such a mechanism, which explains the dynamic behind the masses of fermions and gauge bosons, is still a mystery. There is, moreover, no theoretical reason that the Higgs sector is composed of only one Higgs doublet field. A possible non-minimal Higgs sector with more than one scalar doublet, e.g. the SM with an extra singlet, doublet, and/or triplet, might take place. One of the most important features in models with an extra doublet or triplet is the apparition of charged Higgs bosons, which their discovery would be a clear hint of new physics beyond the SM. 

One of the simplest higher Higgs representations beyond the SM, containing a charged Higgs boson, which merits specific attention, is the Two-Higgs Doublet Model (2HDM). The latter is extended by an additional Higgs doublet in order to generate masses for fermions and gauge bosons. 
The scalar sector of the 2HDM predicts in their spectrum three neural Higgs states
%\footnote{Note that $h$ is, conventionally, predicted to be lighter than $H$ ($M_h < M_H$).}
($h$, $H$ and $A$) and a pair of charged Higgs states ($H^\pm$). One of the two scalars $h$ or $H$ is required to mimic the SM Higgs boson in the so-called \textit{alignment limit} \cite{Ferreira:2012my, Carena:2013ooa, Bernon:2015qea, Bernon:2015wef}. Such a requirement, indeed, puts rather stringent limits on the 2HDM parameter space. Further bounds arise from direct searches at collider experiments and indirect searches from EW precision measurement as well as flavor physics set, moreover, additional bounds on 2HDM.

At the LHC, two conventional ways have been adopted to search for charged Higgs boson, in which both ATLAS and CMS collaborations almost target the fermionic decay channels in their programs to look for $H^\pm$.
When $M_{H^\pm} < m_t - m_b$, a charged Higgs boson has been searching for through the top pair (anti)quark production and decay, $pp \rightarrow t\bar{t} \rightarrow W^\mp H^\pm b \bar{b}$, with $H^\pm$ decaying into a pair of fermions ($\tau\nu$ and $cs$). In case the mass of $H^\pm$ exceeds the mass of top and bottom quarks, the dominant production channel is predicted to be the one associated with $t$ and $b$ ($H^+ tb$ or the equivalent anti-top mode), with $H^+ \rightarrow t b$ or $H^+ \rightarrow \tau^+ \nu$. 
Nevertheless, bosonic decay channels are expected to be naturally dominant over fermionic decays in 2HDM type-I (with branching ratios almost above 99\%) \cite{Akeroyd:1998dt, Enberg:2014pua, Arhrib:2016wpw, Alves:2017snd, Arhrib:2017wmo, Arhrib:2019ywg, Arhrib:2020tqk, Bahl:2021str, Arhrib:2021xmc, Wang:2021pxc, Mondal:2021bxa, Arhrib:2021yqf, Cheung:2022ndq, Abouabid:2022lpg}, i.e. the decay of $H^\pm$ into a $W^\pm$ and a neutral Higgs boson. %Experimentally, such decays have not been the main focus since only a few searches \cite{Wh,WA,WA} are explored so far.
Our study is dedicated to light charged Higgs boson produced via $pp \rightarrow H^\pm h$ followed by $H^\pm \rightarrow W^\pm h$, where $h$ decays into a pair of bottom quarks or tau leptons. The $h \rightarrow \gamma\gamma$ channel will also be investigated here when the $h$ state becomes fermiophobic, in which the coupling of $h$ to fermions is suppressed.

The paper is organized as follows. In Section 2, we introduce the theoretical framework. In Section 3, we present the adopted parameter space scans and applied constraints. In section 4, we discuss our numerical result. We conclude in Section 5.

\section{The 2HDM}
The most popular and motivated extension of the SM, which merits special attention, is the Two-Higgs Doublet Model (2HDM). The latter contains two Higgs doublet fields $\Phi_{1,2}$ with hypercharge $Y_{\Phi_{1,2}}=1$. The 2HDM scalar potential invariant under $SU(2)_L \otimes U(1)_Y$ has the following form \cite{Branco:2011iw}:
\begin{eqnarray}
V(\Phi_1,\Phi_2) &=& m_{11}^2 \Phi_1^\dagger\Phi_1+m_{22}^2\Phi_2^\dagger\Phi_2-[m_{12}^2\Phi_1^\dagger\Phi_2+{\rm h.c.}] \nonumber\\
&+& \frac{1}{2}\lambda_1(\Phi_1^\dagger\Phi_1)^2
+\frac{1}{2}\lambda_2(\Phi_2^\dagger\Phi_2)^2 \nonumber\\
&+&\lambda_3(\Phi_1^\dagger\Phi_1)(\Phi_2^\dagger\Phi_2)
+\lambda_4(\Phi_1^\dagger\Phi_2)(\Phi_2^\dagger\Phi_1) \nonumber\\
&+&\left\{\frac{1}{2}\lambda_5(\Phi_1^\dagger\Phi_2)^2
+\big[\lambda_6(\Phi_1^\dagger\Phi_1)
+\lambda_7(\Phi_2^\dagger\Phi_2)\big]
\Phi_1^\dagger\Phi_2+{\rm h.c.}\right\}, \label{pot1}
\end{eqnarray}
where $m^2_{11}$, $m^2_{22}$ and $\lambda_{1,2,3,4}$ are, in general, real. While $m^2_{12}$ and $\lambda_{5,6,7}$ could be complex to enable CP-violation in the Higgs sector. The natural flavor conservation requirement in the Yukawa sector enforces a $Z_2$ symmetry to the scalar potential. Such a requirement would lead to $m^2_{12}=\lambda_{6,7}=0$. However, one can allow for a soft violation of the discrete symmetry by keeping the $m^2_{12}$ term. In the present study, we shall consider the CP-conserving 2HDM with a softly broken $Z_2$ symmetry, in which $\lambda_{6,7}=0$ and $m^2_{12} \neq 0$.

In the 2HDM, there exist height degrees of freedom. Three of them are absorbed by $W^\pm$ and $Z$ bosons when EWSB occurs. The rest of the degrees of freedom are the five physical Higgs states mentioned above, i.e. two CP-even $h$ and $H$, a CP-odd $A$ and a pair of charged Higgs $H^\pm$.
Under the two minimization conditions, $m^2_{11}$ and $m^2_{22}$ can be substituted by $v_{1,2}$ (which are the vacuum expectation values (vevs) of $\Phi_{1,2}$). Furthermore, $\lambda_{1,2,3,4,5}$ can be traded for $M_{h,H,A,H^\pm}$ and $\sin(\beta-\alpha)$, where $\beta$ and $\alpha$ are the mixing angles. Thus, only seven independent parameters are left: $M_h$, $M_H$, $M_A$, $M_{H^\pm}$, $\sin(\beta-\alpha)$, $\tan\beta$ and $m^2_{12}$. 

In the Yukawa sector, if we enable EWSB similar to the SM (both Higgs doublets couple to fermions) we will end up with FCNCs in such a sector at tree level. The latter can be avoided by forcing the $Z_2$ symmetry such that each fermion type interacts only with one of the Higgs doublets. 
Thus, there are four possible types of Yukawa textures \cite{Branco:2011iw}: type-I, type-II, type-X (or lepton-specific) and type-Y (or flipped). In the present study, we shall focus on type-I 2HDM, which $\Phi_2$ doublet couples to all fermions in order to generate their masses.

The couplings of neutral and charged scalars to fermions can be described by the Yukawa Lagrangian \cite{Branco:2011iw},
\begin{eqnarray}
- {\mathcal{L}}_{\rm Yukawa} = \sum_{f=u,d,l} \left(\frac{m_f}{v} \kappa_f^h \bar{f} f h + 
\frac{m_f}{v}\kappa_f^H \bar{f} f H 
- i \frac{m_f}{v} \kappa_f^A \bar{f} \gamma_5 f A \right) + \nonumber \\
\left(\frac{V_{ud}}{\sqrt{2} v} \bar{u} (m_u \kappa_u^A P_L +
m_d \kappa_d^A P_R) d H^+ + \frac{ m_l \kappa_l^A}{\sqrt{2} v} \bar{\nu}_L l_R H^+ + H.c. \right),
\label{Yukawa-1}
\end{eqnarray}
where $v^2=v^2_1 + v^2_2 \simeq (246 ~\rm{GeV})^2$ and $V_{ud}$ is the CKM matrix element. $P_{L,R}$ refer to left- and right-handed projection operators. $\kappa_f^S$ are the Yukawa couplings, which are given in Table \ref{yuk_coupl} for the four types of 2HDM.
%\begin{center}
\begin{table}[h]
\caption{\label{yuk_coupl} Higgs couplings to fermions for the four types of 2HDM.}
%\footnotesize\rm
\centering
\begin{tabular}{@{}*{10}{c}}
	\br
	 & $\kappa_u^h$ & $\kappa_d^h$ & $\kappa_l^h$ & $\kappa_u^H$ & $\kappa_d^H$ & $\kappa_l^H$ & $\kappa_u^A$ & $\kappa_d^A$ & $\kappa_l^A$  \\
	\mr
	type-I & $c_\alpha/s_\beta$ & $c_\alpha/s_\beta$& $c_\alpha/s_\beta$ & $s_\alpha/s_\beta$ & $s_\alpha/s_\beta$ & $s_\alpha/s_\beta$ & $c_\beta/s_\beta$ & 
	$-c_\beta/s_\beta$ & $-c_\beta/s_\beta$\\
	%\mr
	type-II & $c_\alpha/s_\beta$ & $-s_\alpha/c_\beta$& $-s_\alpha/c_\beta$ & $s_\alpha/s_\beta$ & $c_\alpha/c_\beta$ & $c_\alpha/c_\beta$ & $c_\beta/s_\beta$ & 
	$s_\beta/c_\beta$ & $s_\beta/c_\beta$ \\
	%\mr
	type-X & $c_\alpha/s_\beta$ & $c_\alpha/s_\beta$& $-s_\alpha/c_\beta$ & $s_\alpha/s_\beta$ & $s_\alpha/s_\beta$ & $c_\alpha/c_\beta$ & $c_\beta/s_\beta$ & 
	$-c_\beta/s_\beta$ & $s_\beta/c_\beta$ \\ 
	type-Y & $c_\alpha/s_\beta$ & $-s_\alpha/c_\beta$& $c_\alpha/s_\beta$ & $s_\alpha/s_\beta$ & $c_\alpha/c_\beta$ & $s_\alpha/s_\beta$ & $c_\beta/s_\beta$ & 
	$s_\beta/c_\beta$ & $-c_\beta/s_\beta$ \\ 
	\br
\end{tabular}
\end{table}
%\end{center}

\section{2HDM scans and constraints}
We perform a random scan over the 2HDM parameter space using the program \texttt{2HDMC-1.8.0} \cite{Eriksson:2009ws}. While we assume that $H$ is the observed SM-like Higgs fixing its mass to $125.09~\rm{GeV}$, we demand $M_h$ to lie in the $15$ and $120~\rm{GeV}$ range. We further scan $M_A$ within the range between $15$ and $200~\rm{GeV}$, $M_{H^\pm}$ between $80$ and $170$, $\sin(\beta-\alpha)$ between $-0.5$ and $0.5$, $\tan\beta$ between $2$ and $50$, and $m^2_{12}$ between $0$ and $m^2_h \sin\beta \cos\beta$. During the scan, we require that each parameter point complies with the following theoretical and experimental constraints: 
\begin{itemize}
	\item Unitarity, perturbativity and vaccum stability using the program \texttt{2HDMC}.  
	\item The parameters $S$, $T$ and $U$, which constrain the mass splitting between the physical scalars, using the result ($U=0$) of Ref. \cite{Haller:2018nnx}. 
	
	\item Limits from searches for additional Higgs bosons using the program \texttt{HiggsBounds-5.9.0} \cite{Bechtle:2020pkv}. 
	
	\item Compliance with SM-like Higgs boson measurements using the program \texttt{HiggsSignals-2.6.0} \cite{Bechtle:2020uwn}.
	
	\item Compliance with the $Z$ width measurement from LEP, $\Gamma_Z = 2.4952 \pm 0.0023$ GeV \cite{ALEPH:2005ab}, in which $\Gamma(Z \rightarrow h A)$ is taken to be within the $2\sigma$ experimental uncertainty of such a measurement.
	
	\item Flavor physics constraints using the program \texttt{SuperIso v4.1} \cite{Mahmoudi:2008tp}. The experimental results of relevant observables are taken from Ref. \cite{Haller:2018nnx}.
\end{itemize}

In order to satisfy the above constraints, especially those from electroweak oblique parameters $S$ and $T$, the Higgs states should be relatively close in mass. Note that, after considering the new CDF W boson mass measurement \cite{CDF:2022hxs}, it is found recently that the charged Higgs boson is always heavier than the neutral Higgs states in 2HDM with inverted hierarchy ($H \equiv H_{\mathrm{SM}}$) \cite{Abouabid:2022lpg}.

\section{Results and discussion}
As mentioned above, we focus on light charged Higgs boson in  $H^\pm h$ associated production channel. We calculate the cross section at the leading order with $\sqrt{s} = 14~\rm{TeV}$ using the code \texttt{MadGraph5\_aMC@NLO-2.7.3} \cite{Alwall:2014hca}. In the calculation, we have used the MMHT2014 pdf set \cite{Harland-Lang:2014zoa}. Note that the NNLO-QCD corrections increase the tree-level cross section by a factor of approximately $1.35$ \cite{Bahl:2021str}. In our study, we shall present the tree-level results only.

\begin{figure}[H]
	\centering
	\begin{subfigure}{0.42\textwidth}
        \includegraphics[width=17pc]{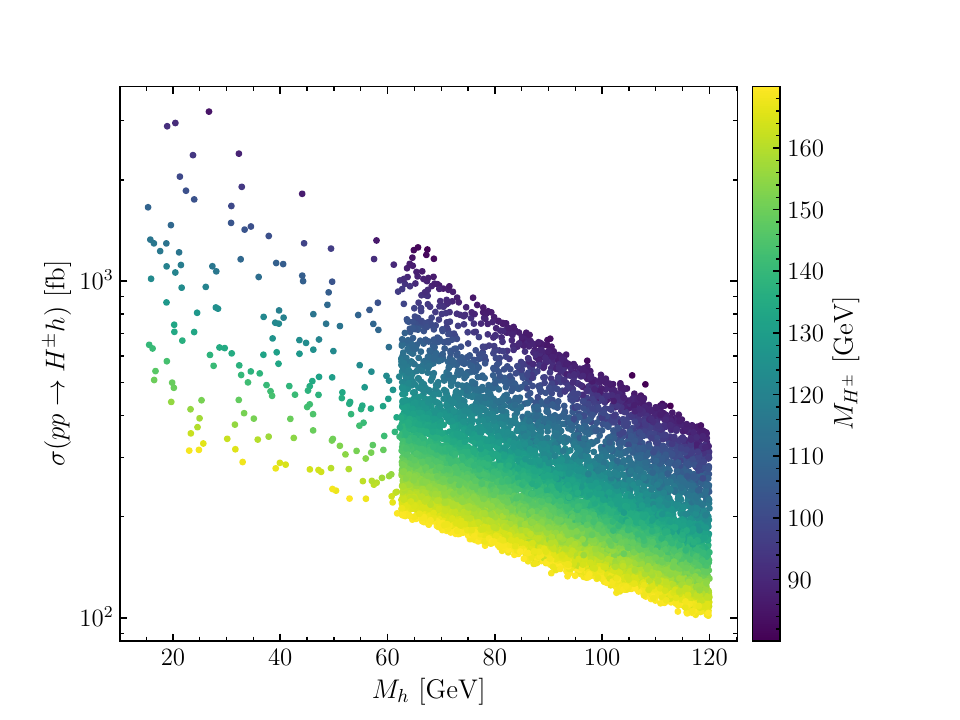}
		\caption{} \label{fig:1a}
	\end{subfigure}%
    \begin{subfigure}{0.42\textwidth}
		\includegraphics[width=17pc]{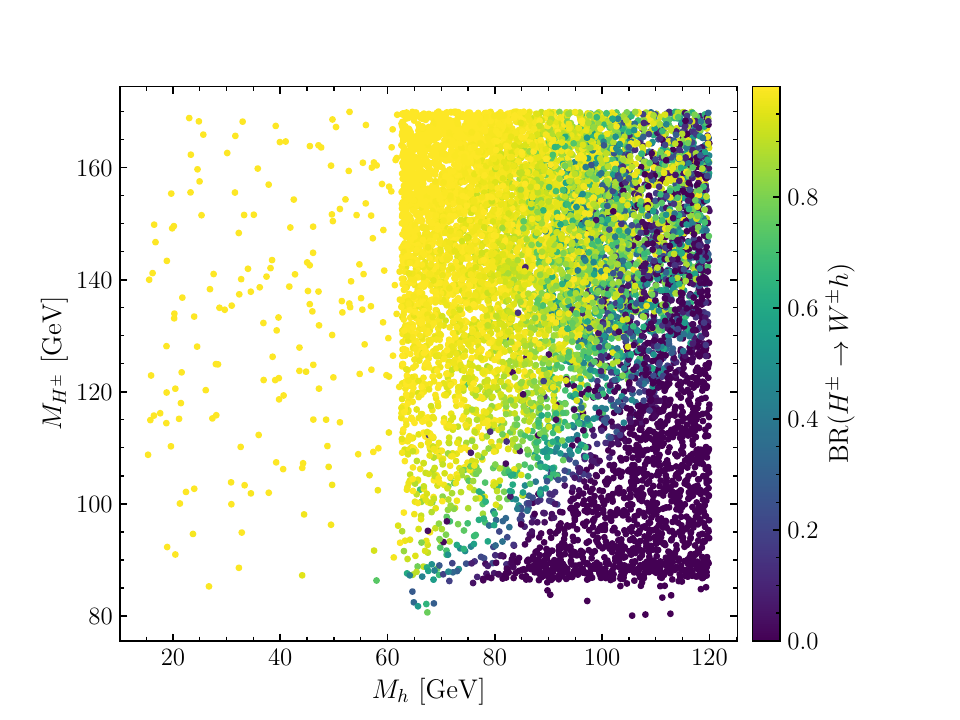}
	    \caption{} \label{fig:1b}
    \end{subfigure}%
	    %\centering
		%\includegraphics[width=16pc]{figs/xs_ppHch.pdf}
		%\includegraphics[width=16pc]{figs/BRHpWh.pdf}
		\caption{\label{Figure1} (a) Cross section for $pp \rightarrow H^\pm h$ at $\sqrt{s} = 14~\rm{TeV}$ as a function of $M_h$, with the color bar indicates $M_{H^\pm}$. (b) Branching ratio BR$(H^\pm \rightarrow W^\pm h)$ in the ($M_h$, $M_{H^\pm}$) plane.}
\end{figure}

In Figure \ref{fig:1a}, we show the values of $\sigma(pp \rightarrow H^\pm h)$ as a function of $M_h$, with the color code indicates $M_{H^\pm}$, for the parameter points that survive all constraints. It is clearly visible that the production cross section could reach the pb level for small Higgs masses. The results can be, indeed, interpreted by the two factors: the $H^\pm W^\mp h$ vertex, which is proportional to $\cos(\beta-\alpha)$ and is maximized in the \textit{alignment limit}; the phase space attributable to relatively small values of $M_h$ and $M_{H^\pm}$. The cross section for $pp \rightarrow H^\pm h$ can, in
certain scenarios, exceed the light charged Higgs production through top quark decay channel, i.e. $pp \rightarrow \bar{t}t \rightarrow \bar{t}bH^+$, especially when $\tan\beta$ is large in which the $H^\pm$ couplings to fermions is suppressed in 2HDM type-I.
 
Let us now turn to the charged Higgs decaying into a $W^\pm$ boson and an $h$ boson, $H^\pm \rightarrow W^\pm h$. Analogous to the $pp \rightarrow H^\pm h$ production cross section, the decay $H^\pm \rightarrow W^\pm h$ is governed by the same vertex and also is maximized in the \textit{alignment limit}. Figure \ref{fig:1b} clearly demonstrates that the branching ratio BR$(H^\pm \rightarrow W^\pm h)$ can reach $100\%$ for light $h$. Note that BR$(H^\pm \rightarrow W^\pm A)$ is only relevant when $M_A \lesssim 100~\mathrm{GeV}$. BR$(H^\pm \rightarrow W^\pm H)$ is suppressed in our scenario by the factor $\sin^2(\beta-\alpha)$. 

\begin{figure}[H]
	\centering
	\begin{subfigure}{0.33\textwidth}
		\includegraphics[width=14pc]{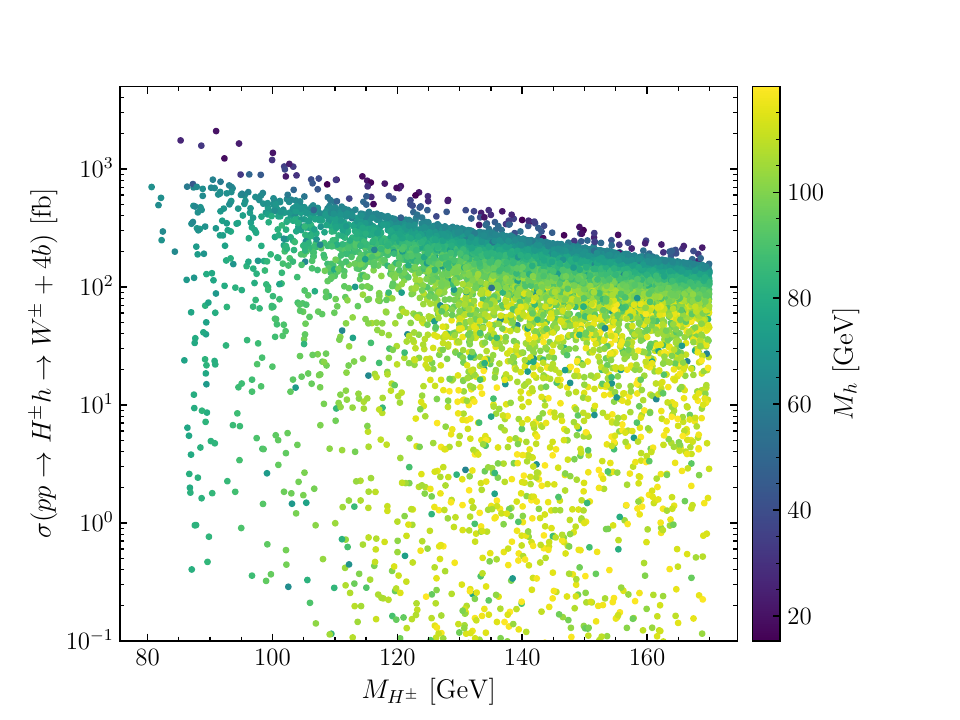}
		\caption{} \label{fig:2a}
	\end{subfigure}%
	\begin{subfigure}{0.33\textwidth}
		\includegraphics[width=14pc]{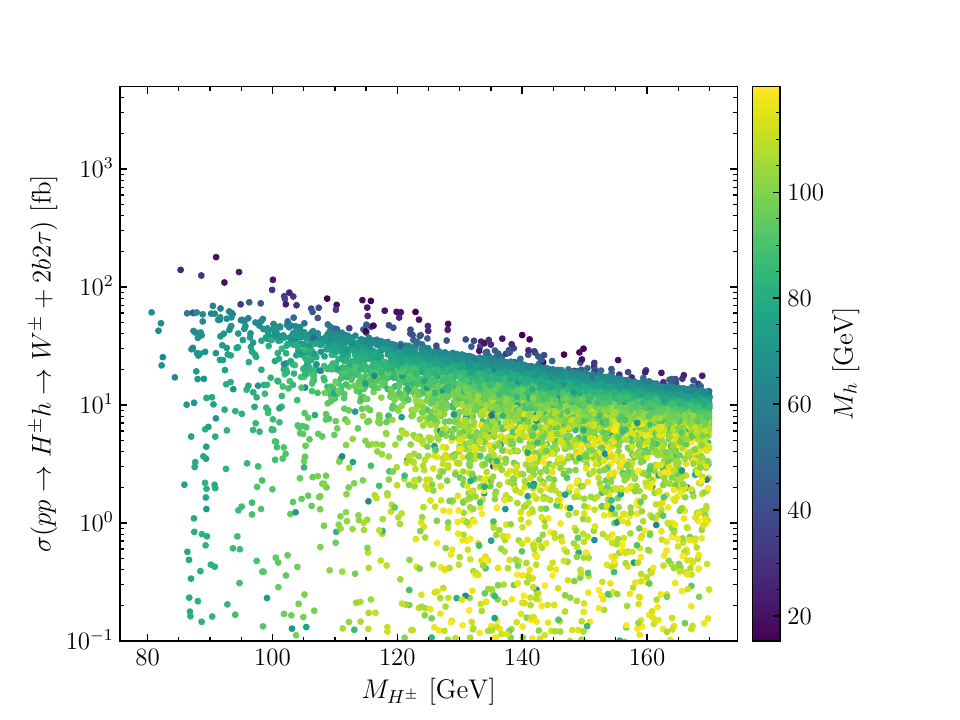}
		\caption{} \label{fig:2b}
	\end{subfigure}%
	\begin{subfigure}{0.33\textwidth}
		\includegraphics[width=14pc]{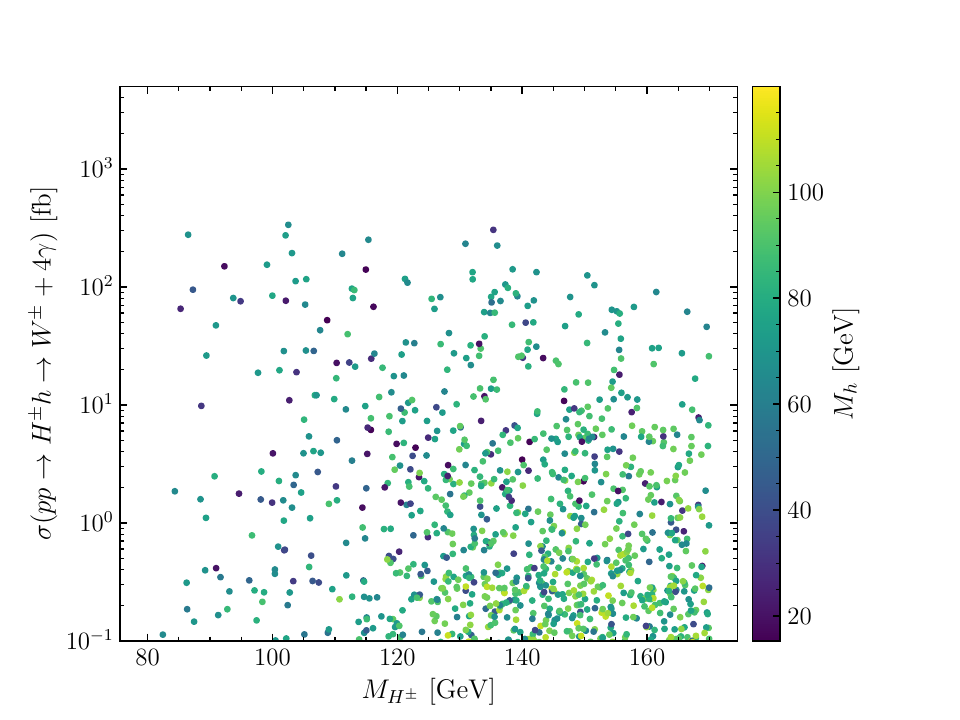}
		\caption{} \label{fig:2c}
	\end{subfigure}%
	%\centering
	%\includegraphics[width=16pc]{figs/xs_ppHch.pdf}
	%\includegraphics[width=16pc]{figs/BRHpWh.pdf}
	\caption{\label{Figure2} (a) Signal cross sections $\sigma(pp \rightarrow H^\pm h \rightarrow W^\pm h h \rightarrow W^\pm + 4b)$, (b) $\sigma(pp \rightarrow H^\pm h \rightarrow W^\pm h h \rightarrow W^\pm + 2b2\tau)$ and (c) $\sigma(pp \rightarrow H^\pm h \rightarrow W^\pm h h \rightarrow W^\pm + 4\gamma)$ as a function of $M_{H^\pm}$, with the color map showing $M_{h}$.}
\end{figure}

Under these observations, combining the $H^\pm h$ production and the $H^\pm \rightarrow W^\pm h$ decay, one can expect significant signatures, which might be alternative discovery channels for light charged Higgs boson at the LHC \cite{Arhrib:2021xmc, Wang:2021pxc, Arhrib:2021yqf}. For the neutral Higgs boson, $h$, we mainly consider $bb$, $\tau\tau$ and $\gamma\gamma$ decays. These channels would lead to $W+4b$, $W+2b2\tau$ and $W+4\gamma$ final states, which are depicted in Figure \ref{Figure2}. One can see (Figure \ref{fig:2a}) that the signal cross section $\sigma(pp \rightarrow H^\pm h \rightarrow W^\pm h h \rightarrow W^\pm + 4b)$ can reach the pb level for very light $h$. Similar to the $W+4b$ final state, in figure \ref{fig:2b} one sees that  $\sigma(pp \rightarrow H^\pm h \rightarrow W^\pm h h \rightarrow W^\pm + 2b2\tau)$ can reach up to $200~\mathrm{fb}$ for low $M_h$. Besides $W+4b$ and $W+2b2\tau$ final states, and since the $h \rightarrow \gamma\gamma$ decay is experimentally the cleanest decay channel (also $WW$ and $ZZ$ decays), which is dominant (close to the \textit{fermiophobic limit}) before the $WW^*$ threshold, the $W+4\gamma$ signature could be an excellent experimental avenue to identify a light charged Higgs boson at the LHC (Figure \ref{fig:2c}). Furthermore, it is found that the signal $pp \rightarrow H^\pm h \rightarrow W^\pm h h \rightarrow W^\pm + 4\gamma$ (assuming $W^\pm \rightarrow \ell^\pm \nu$) is essentially background free, yielding to a large significance \cite{Wang:2021pxc}.

\section{Conclusion}
In this study, we have investigated the light charged Higgs boson ($H^\pm$) production in association with a neutral Higgs boson ($h$) in a scenario where $H$ is identified to be the SM-like state of mass $125~\mathrm{GeV}$, and the parameter space consistent with theoretical and experimental constraints. Under these conditions, we have mainly focused on the $W^\pm h$ decay channel of $H^\pm$,  which is naturally dominant close to the \textit{alignment limit}. We have demonstrated that such production and decay channels would lead to significant cross sections for $W^\pm + 4b$, $W^\pm + 2b2\tau$ and $W^\pm + 4\gamma$ final states, which might be alternative signatures to target the light charged Higgs boson at the LHC.  
  
\section*{Acknowledgments}
The work of AA, RB and BM is supported by the Moroccan Ministry of Higher Education and Scientific Research MESRSFC and CNRST Project PPR/2015/6. The work of
SM is supported in part through the NExT Institute and the STFC Consolidated Grant No. ST/L000296/1. 
Y. W. is supported by the Scientific Research Funding Project for Introduced High-level Talents of the Inner Mongolia Normal University Grant No. 2019YJRC001, and the scientific research funding for introduced high-level talents of Inner Mongolia of China. 
Q.-S. Yan work is supported by the Natural Science Foundation of China Grant No. 11875260.
%\newpage
\section*{References}
\bibliography{PanAfr-Krab}

\providecommand{\newblock}{}
\begin{thebibliography}{10}
\expandafter\ifx\csname url\endcsname\relax
  \def\url#1{{\tt #1}}\fi
\expandafter\ifx\csname urlprefix\endcsname\relax\def\urlprefix{URL }\fi
\providecommand{\eprint}[2][]{\url{#2}}

\bibitem{ATLAS:2012yve}
Aad G {\em et~al.\/} (ATLAS) 2012 {\em Phys. Lett. B\/} {\bf 716} 1--29
  (\textit{Preprint} \eprint{1207.7214})

\bibitem{CMS:2012qbp}
Chatrchyan S {\em et~al.\/} (CMS) 2012 {\em Phys. Lett. B\/} {\bf 716} 30--61
  (\textit{Preprint} \eprint{1207.7235})

\bibitem{Ferreira:2012my}
Ferreira P~M, Santos R, Sher M and Silva J~P 2012 {\em Phys. Rev. D\/} {\bf 85}
  035020 (\textit{Preprint} \eprint{1201.0019})

\bibitem{Carena:2013ooa}
Carena M, Low I, Shah N~R and Wagner C~E~M 2014 {\em JHEP\/} {\bf 04} 015
  (\textit{Preprint} \eprint{1310.2248})

\bibitem{Bernon:2015qea}
Bernon J, Gunion J~F, Haber H~E, Jiang Y and Kraml S 2015 {\em Phys. Rev. D\/}
  {\bf 92} 075004 (\textit{Preprint} \eprint{1507.00933})

\bibitem{Bernon:2015wef}
Bernon J, Gunion J~F, Haber H~E, Jiang Y and Kraml S 2016 {\em Phys. Rev. D\/}
  {\bf 93} 035027 (\textit{Preprint} \eprint{1511.03682})

\bibitem{Akeroyd:1998dt}
Akeroyd A~G 1999 {\em Nucl. Phys. B\/} {\bf 544} 557--575 (\textit{Preprint}
  \eprint{hep-ph/9806337})

\bibitem{Enberg:2014pua}
Enberg R, Klemm W, Moretti S, Munir S and Wouda G 2015 {\em Nucl. Phys. B\/}
  {\bf 893} 420--442 (\textit{Preprint} \eprint{1412.5814})

\bibitem{Arhrib:2016wpw}
Arhrib A, Benbrik R and Moretti S 2017 {\em Eur. Phys. J. C\/} {\bf 77} 621
  (\textit{Preprint} \eprint{1607.02402})

\bibitem{Alves:2017snd}
Alves D~S~M, El~Hedri S, Taki A~M and Weiner N 2017 {\em Phys. Rev. D\/} {\bf
  96} 075032 (\textit{Preprint} \eprint{1703.06834})

\bibitem{Arhrib:2017wmo}
Arhrib A, Benbrik R, Enberg R, Klemm W, Moretti S and Munir S 2017 {\em Phys.
  Lett. B\/} {\bf 774} 591--598 (\textit{Preprint} \eprint{1706.01964})

\bibitem{Arhrib:2019ywg}
Arhrib A, Cheung K and Lu C~T 2020 {\em Phys. Rev. D\/} {\bf 102} 095026
  (\textit{Preprint} \eprint{1910.02571})

\bibitem{Arhrib:2020tqk}
Arhrib A, Benbrik R, Harouiz H, Moretti S, Wang Y and Yan Q~S 2020 {\em Phys.
  Rev. D\/} {\bf 102} 115040 (\textit{Preprint} \eprint{2003.11108})

\bibitem{Bahl:2021str}
Bahl H, Stefaniak T and Wittbrodt J 2021 {\em JHEP\/} {\bf 06} 183
  (\textit{Preprint} \eprint{2103.07484})

\bibitem{Arhrib:2021xmc}
Arhrib A, Benbrik R, Krab M, Manaut B, Moretti S, Wang Y and Yan Q~S 2021 {\em
  JHEP\/} {\bf 10} 073 (\textit{Preprint} \eprint{2106.13656})

\bibitem{Wang:2021pxc}
Wang Y, Arhrib A, Benbrik R, Krab M, Manaut B, Moretti S and Yan Q~S 2021 {\em
  JHEP\/} {\bf 12} 021 (\textit{Preprint} \eprint{2107.01451})

\bibitem{Mondal:2021bxa}
Mondal T and Sanyal P 2022 {\em JHEP\/} {\bf 05} 040 (\textit{Preprint}
  \eprint{2109.05682})

\bibitem{Arhrib:2021yqf}
Arhrib A, Benbrik R, Krab M, Manaut B, Moretti S, Wang Y and Yan Q~S 2021 {\em
  Symmetry\/} {\bf 13} 2319 (\textit{Preprint} \eprint{2110.04823})

\bibitem{Cheung:2022ndq}
Cheung K, Jueid A, Kim J, Lee S, Lu C~T and Song J 2022  (\textit{Preprint}
  \eprint{2201.06890})

\bibitem{Abouabid:2022lpg}
Abouabid H, Arhrib A, Benbrik R, Krab M and Ouchemhou M 2022
  (\textit{Preprint} \eprint{2204.12018})

\bibitem{Branco:2011iw}
Branco G~C, Ferreira P~M, Lavoura L, Rebelo M~N, Sher M and Silva J~P 2012 {\em
  Phys. Rept.\/} {\bf 516} 1--102 (\textit{Preprint} \eprint{1106.0034})

\bibitem{Eriksson:2009ws}
Eriksson D, Rathsman J and Stal O 2010 {\em Comput. Phys. Commun.\/} {\bf 181}
  189--205 (\textit{Preprint} \eprint{0902.0851})

\bibitem{Haller:2018nnx}
Haller J, Hoecker A, Kogler R, M\"onig K, Peiffer T and Stelzer J 2018 {\em
  Eur. Phys. J. C\/} {\bf 78} 675 (\textit{Preprint} \eprint{1803.01853})

\bibitem{Bechtle:2020pkv}
Bechtle P, Dercks D, Heinemeyer S, Klingl T, Stefaniak T, Weiglein G and
  Wittbrodt J 2020 {\em Eur. Phys. J. C\/} {\bf 80} 1211 (\textit{Preprint}
  \eprint{2006.06007})

\bibitem{Bechtle:2020uwn}
Bechtle P, Heinemeyer S, Klingl T, Stefaniak T, Weiglein G and Wittbrodt J 2021
  {\em Eur. Phys. J. C\/} {\bf 81} 145 (\textit{Preprint} \eprint{2012.09197})

\bibitem{ALEPH:2005ab}
Schael S {\em et~al.\/} (ALEPH, DELPHI, L3, OPAL, SLD, LEP Electroweak Working
  Group, SLD Electroweak Group, SLD Heavy Flavour Group) 2006 {\em Phys.
  Rept.\/} {\bf 427} 257--454 (\textit{Preprint} \eprint{hep-ex/0509008})

\bibitem{Mahmoudi:2008tp}
Mahmoudi F 2009 {\em Comput. Phys. Commun.\/} {\bf 180} 1579--1613
  (\textit{Preprint} \eprint{0808.3144})

\bibitem{CDF:2022hxs}
Aaltonen T {\em et~al.\/} (CDF) 2022 {\em Science\/} {\bf 376} 170--176

\bibitem{Alwall:2014hca}
Alwall J, Frederix R, Frixione S, Hirschi V, Maltoni F, Mattelaer O, Shao H~S,
  Stelzer T, Torrielli P and Zaro M 2014 {\em JHEP\/} {\bf 07} 079
  (\textit{Preprint} \eprint{1405.0301})

\bibitem{Harland-Lang:2014zoa}
Harland-Lang L~A, Martin A~D, Motylinski P and Thorne R~S 2015 {\em Eur. Phys.
  J. C\/} {\bf 75} 204 (\textit{Preprint} \eprint{1412.3989})

\end{thebibliography}

\end{document}